\documentclass{article}

\begin{document}
\newcommand*{\availability}{availability}	
\newcommand*{\availabilities}{availabilities}	

\title{Clocks And Dynamics In Quantum Mechanics}         
\author{York, Michael\footnote{michael.york@physics.oxon.org}}        
\date{\today}          
\maketitle

\abstract{We argue that (1) our perception of time through change and (2) the gap between reality and our observation of it are at the heart of both quantum mechanics and the dynamical mechanism of physical systems.  We suggest that the origin of quantum uncertainty lies with the absence of infinities or infinitessimals in observational data and that our concept of time derives from observing changing data (events). We argue that the fundamentally important content of the Superposition Principle is not the ``probability amplitude'' of {\em posterior} state observation but future state {\em \availability} conditional only on {\em prior} information. Since event detection also implies posterior conditions (e.g. a specific {\em type} of detectable event occurred) as well as prior conditions, the probabilities of detected outcomes are also conditional on properties of the posterior properties of the observation. Such posterior conditions cannot affect the prior state \availabilities\ and this implies violation of {\em counter-factual definiteness}. 

A component of a quantum system may be chosen to represent a clock and changes in other components can then be expected to be correlated with clocks with which they are entangled. Instead of traditional time-dependent equations of motion we provide a specific mechanism whereby evolution of data is instead quasi-causally related to the relative \availability\ of states and equations of motion are expressed in terms of quantized clock variables. We also suggest that time-reversal symmetry-breaking in weak interactions is an artifice of a conventional choice of co-ordinate time-function. Analysis of a ``free'' particle suggests that conventional co-ordinate space-time emerges from how we measure the separation of objects and events.}

\section{Introduction}\label{sec:intro}      

Since the introduction of quantum mechanics (QM) into theoretical physics at the beginning of the 20th century, a debate has raged -- and is still raging -- over what it means. Of the mathematical formalism, there can be little doubt; the experimental evidence is that it works extremely well in every area that it has been tested. What is more, an astoundingly wide-ranging relativistic quantum field theory (RQFT) of how the fundamental constituents of matter interact has been built on top of it in parallel to a string theory that has its origins in the S-matrix program. This is not to claim that either of these theories is complete and irrefutable. But RQFT in particular has given us an encyclopedia of successful calculation of observable quantities. Whether or not the mathematical formalism of QM is complete or not, it is clear that it represents a major advance over pre-QM theory.

Perhaps the most currently difficult problem lies with gravity. Although RQFT has been successful in integrating every other known force, gravity still eludes it. String theory has made proposals for such unification but they also come accompanied by problems and are untested. Modern understanding of gravity is oriented around the general relativistic theory  of space-time geometry. We conjecture that perhaps a successful integration of gravity with QM may lie first of all with a re-examination of the concept of a continuous time co-ordinate in QM. As an effort in this direction, it is the purpose of this paper to propose a framework in which interaction implies the presence of objects which serve as clocks and -- depending on the choice of clocks -- with which the rest of the physical system may be entangled. We shall also see that a corollary of the framework we present is that our concept of a continuous spatial co-ordinate frame also needs re-examination.

Despite numerous attempts, the ``interpretation'' of QM in terms that the pre-QM scientist might understand is something that has never been resolved. In fact, in the opinion of this author, it cannot be resolved for the fundamental reason that it relies on an inappropriate concept of time. The historical development of QM followed the time-dependent path of wave equations and then field theory. Yet the classical concept of co-ordinate time employed, even in the relativistic case, stands on very shaky ground. 

On the other hand time-independent QM -- as employed in S-matrix theory, for instance -- has both a philosophical basis that is more appealing (to this author at least) and is thoroughly tested in particle collisions. One of the interesting aspects of this is that no outside clock is necessary to describe particle collisions; the collision itself is the tick on a clock and all of the physics is in the relative probabilities of the outcomes. However, its theoretical efficacy in the past has been mostly limited to strong interactions. It is our suggestion here that real understanding of QM involves new principles that are completely outside the classical framework of a continuous time co-ordinate, such as those of S-matrix theory.

These new principles - which we discuss in more detail in section \ref{sec:fundamentals} -- are all related to the discretization of observational data including clock data, the possible entanglement of clocks with the system under observation, the relative similarity of posterior states to prior states, the selectivity of the observational context and conservation principles.

In section \ref{sec:Hilbert} we tie these principles together in a more specific mathematical way in the Hilbert space picture of observable states, introduce the concept of state {\em availability}, show the differences from the conventional interpretation of QM, specify the general equations of observable motion in terms of clock variables and see how this puts our conventional view of space-time into question. In section \ref{sec:compute} we provide a general method for computing availabilities and in section \ref{sec:chooseClocks} we discuss the choice of clocks and co-ordinate time-function.


\section{Fundamentals}\label{sec:fundamentals}

In the early days of QM, a possible conflict between local reality and QM was highlighted by Einstein, Podolsky and Rosen \cite{EPR}. More recently, the theoretical work of Bell \cite{Bell} and experiments of Clauser, Aspect and others \cite{C1},\cite{C2},\cite{C3},\cite{A1},\cite{A2},\cite{A3} have shown that unless we give up {\em locality} or allow nature to conspire to fool us, then reality violates {\em counterfactual definiteness}\footnote{This implies limitations on the validity of projecting implied observation (based on theoretical physical laws) from actual observation.}. In this section we shall explain why we interpret this violation as meaning that a superposition is not a {\em physical} property of reality between observations, but merely of our picture of reality which makes it a predictive tool for the outcome probability of an {\em actual} observation {\em only if it were made} and link it to the discrete nature of observational data. 

In fact it will be a fundamental feature of the framework we present here that QM is not a property of {\em reality} itself but a key essential property of the inevitable gap between reality and our necessarily limited picture of it and this is the key non-classical ingredient we consider necessary to understanding QM and banishing all the so-called paradoxes.

The need for a framework that acknowledges this gap can be understood in the context of centuries of philosophical debate between the two extremes of the ultra-realist school and the ultra-idealist school. According to the former, we observe reality objectively. But for the latter, reality is a subjective construct of our dreams. Of course, there exist many schools of thought between these two. We will not discuss them any further, but simply lay out our own framework for our collective picture of nature on the assumption that reality is an inherent assumption of our collective attempts at constructing logically consistent theories of it and the project we call physics is focused on making that collective picture as objective\footnote{In the sense of not being dependent on the differing qualities or experiences of the humans building it.} as possible.

There are nine fundamental principles to this framework which we shall state in general terms as:

\begin{enumerate}
  \item {{\em Finite Measurement}: All measured (observable) data must be finite in value.}\label{prin:measurement}
  \item {{\em Discrete Transition}: Things change and no measured change may be infinitesimal.}\label{prin:discrete}
  \item {{\em Entanglement}: Changes in some objects are correlated with changes in others.}\label{prin:entangle}
  \item {{\em Chronology}: We may choose a changing observable to be a clock. A clock then serves as a reference relative to which we measure change in other observables.}\label{prin:chrono}
  \item {{\em Selectivity}: Observation requires or implies selection of observable states -- including clocks.}\label{prin:selectivity}
  \item {{\em Symmetry}: Given an observational environment, observed change is determined by the symmetry properties implicit in our picture of reality and these properties determine a state space that is independent of the observer and the observational selections made.}\label{prin:symmetry}
  \item {{\em State Proximity}: For small changes on a clock, small changes in other entangled observables are more likely than large changes.}\label{prin:proximity}
  \item {{\em Probabilistic Dynamics}: The probability of observing a specific state, given a prior state, is dependent on the physical availability of the final state conditional on the prior state.} \label{prin:prob}
  \item {{\em Isolation And Conservation}: Any system that remains isolated between observations will conserve certain external composite properties although its internal structure may otherwise change without restriction.}\label{prop:conserve}
\end{enumerate}

We'll assume that principle \ref{prin:measurement} is self-evident and not discuss it further. 

Principle \ref{prin:discrete} says that although we may, in principle, detect arbitrarily small changes, they must always be finite. Since no detectable change can be infinitesimal, all data (from our collective observation and communication) comes to us in discrete form; {\em we cannot observe continuous change}. 

Principles \ref{prin:entangle} and \ref{prin:chrono} concern our concept of time and the ``dynamics'' which relate changes in some objects to changes in others. We impute the meaningfulness of our idea of time from our observation of change and the tendency of things to break up (lose structure) rather than to come together (gain structure) -- which gives us a sense of past and future. We measure it in terms of changes in observables we call clocks by defining a specific monotonic analytic function $t(c)$ where $c$ is the measured property (or set of properties) of any chosen clock. The implication of our principle \ref{prin:chrono}, however, is that it is the clock observables $c$ that are sufficient to determine the dynamics of a system and that the time-function $t(c)$ is unnecessary except as a means to find a co-ordinate by which to measure time in a way common to all clocks (at least for a given observer).

From principles \ref{prin:measurement} - \ref{prin:chrono} we can deduce that there is no continuously observable time-function $t(c)$. This does not mean that we cannot, under appropriate circumstances, conceive a continuous trajectory of an object by plotting successive positions against successive clock times and compute a ``velocity'' by taking the ratio of small changes in the observable and $t(c)$. But it does mean that, in this case, the velocity is a property of the observable transition rather than the observable object\footnote{It also means that a trajectory may change or even break down completely if an object interacts with another object in the interval between observations and is deflected or transmuted. So we can't even be sure we are seeing the {\em same} object with each observation {\em even if we see an object with the same properties}.}. In particular, the impossibility of infinitesimal change means that we cannot simultaneously measure an observable and its rate of change and, in general, we see that the impossibility of the simultaneous observation of incompatible observables is a necessary consequence. Unobservability between the arrival of data packets imposes a necessary and fundamental limitation on our ability to relate our picture of reality to reality itself. We can also deduce from our first four principles that, since any change in any observable implies there must be an observable change in any reasonable choice of clock, then the rate of change of  any observable must be finite and thus, in particular, no object can carry information instantaneously across space\footnote{Entanglement, on the other hand, does not depend on change in a clock variable or a time-function and may persist however far apart entangled objects may be.}.

Principle \ref{prin:selectivity} is the statement that the design of any experiment conditions the outcome. This can be in setting up an initial state as well as in determining what later states can be detected. Even unplanned event detection is conditional on the detection environment just as human observation is conditional on our biological apparatus. In general, in any given experiment, posterior conditions on the outcome exist as well as prior conditions on the set-up and this has implications for the ``role of the observer''. In particular, it means that the space of {\em detectable} outcomes will usually be smaller than the space of {\em possible} outcomes as well as dependent on the observer's frame of reference. At a minimum, the occurrence of an event itself is a posterior condition for there to be an outcome. Although it may seem trivial, an observation is a posterior condition of an observed outcome. Without such a condition we have no outcome. But, absent prescience on the part of our pictured real systems, the supposed state of our pictured reality between the set-up and the outcome cannot itself be conditional on the -- as yet undetermined -- conditions of outcome detection. In particular, this is an example of how {\em counterfactual definiteness} is violated by the posterior conditions of detection.

Principle \ref{prin:symmetry} is our equivalent to the invariance principles of the laws of physics combined with a recognition that those symmetry principles determine the state space and what we see as ``time-dependence'' depends on our choice of clock and time function. For example, whatever choice of clock we make, we could require that the choice of $t(c)$ ensures compatibility with Lorentz invariant equations of motion -- if we believe that invariance principle to hold. 

Principle \ref{prin:proximity} is very similar to -- but should be seen as more general than -- conventional {\em local causality} (in which observable change must be time-like). It is not a requirement that an observed state be similar to a previous observed state. It is rather a requirement for making a meaningful connection between successive observations that appear to have some connection between (e.g. approximate continuity in) their observable properties.

Principle \ref{prin:prob} can be recognized as somewhat similar to the conventional probability assumption in QM. However, it differs from the conventional framework of ``probability amplitudes'' by dealing with state \availabilities\ -- a subtlety that we shall develop further in the next section. We naturally expect his dependence of probability on \availability\ to be monotonic (the more available a state, the more probable), but it will also be subject to the selectivity principle \ref{prin:selectivity}.

Note that principles \ref{prin:chrono} and \ref{prin:prob} together imply that the  apparent time-dependence (dynamical evolution) of a physical system is a construction obtained from the entanglement of a set of observables with a separate clock observable and the resulting correlation of their changes with changes in clock states.

Principle \ref{prop:conserve}, although apparently uncontroversial, will be the cornerstone of our principle departure from conventional QM. In particular because, when taken together with the other properties described here, it means that only the composite state of an isolated system is preserved. Any state we think might describe the internal properties is unknowable between observations -- even if we knew it at a prior observation -- and the posterior observation may even occur in a completely different subsystem space to that of the prior observation as long as the conservation rule is obeyed for the external properties of the composite system. This is how transitions occur in our framework and will be our alternative to the conventional time-dependent evolution of a system according to a pre-ordained Hamiltonian. This conservation rule then becomes the most we can say about reality itself between observations -- and then only for an isolated system.

\section{The State Space}\label{sec:Hilbert}

As usual, we assume that the space of possible states that, in principle at least, could be observed can be represented by vectors in a Hilbert space.

We then define the state vector $|q>$ to represent the system where $q$ is a description of any possible observable state (e.g. a collection of potentially observable quantum numbers in a given frame of reference) and then $|q>$ is a state vector in the global space. 

However, we part from convention here by not assigning time to be a separately measurable quantity. Rather let us assume that our system consists of two subsystems, and that we can choose one such subsystem to represent a clock. Then we can write
\begin{eqnarray}
\label{eqn:chooseClock}|q> & = & |r,c>
\end{eqnarray} 
where $c$ specifies the observable features of our clock and $r$ those of the remaining subsystem. Then the state vector $|r,c>$ also exists in the direct product space of the clock and remainder. Clearly, $c$ and $r$ {\em may}, in general, be {\em entangled} in the global space and it is this entanglement that generates the dynamics of the remainder subsystem with respect to the clock. 

The composite state space of an isolated system with conserved quantum numbers $Q$ will obey the conventional orthogonality and completeness conditions as:
\begin{eqnarray}
\label{eqn:orthogonal}<Q'|Q> & = & \delta_{Q'Q}\\
\label{eqn:complete}\sum_{Q} |Q><Q| & = & 1
\end{eqnarray} 

We can always write the observed state, whatever its internal structure, as a superposition within the composite Hilbert space. For the prior state vector this is
\begin{eqnarray}
\label{eqn:qSuperpos}|q> & = & \sum_{Q} |Q><Q|q>
\end{eqnarray} 
and similarly for the posterior state vector. We shall define the posterior \availability\ of state $q'$ given a prior state $q$ to be the magnitude of the scalar product of their state vectors $|<q'|q>|$ in the composite\footnote{It is important to understand that the scalar product is taken in the composite space and not in the component space. In the latter case, the scalar product would always be trivially zero or unity.} space of all observable states. We then see how condition \ref{eqn:orthogonal} entails the conservation of $Q$, since $Q'=Q$ is the only available final state in the $Q$-space of an isolated system given initial $Q$.  

Clearly, when $q$ and $q'$ are similar we expect $|q'>$ and $|q>$ to be close and the \availability\ to be larger than when they are very different. The completeness relation \ref{eqn:complete}, enables us to write 
\begin{eqnarray}
\label{eqn:qScalarProd} <q'|q> & = & \sum_{Q} <q'|Q><Q|q>
\end{eqnarray} 
showing that state \availability\ can be calculated in terms of the instantaneous vector-coupling coefficients for the observed state -- as we shall see in more detail in section \ref{sec:compute}. Note that in the above expression the left hand coupling of each $|Q>$ is with the posterior component product space and the right hand coupling with the prior component product space. 

Although the reason we call the \availability\ by that name is fairly transparent, the precise physical meaning has not yet been specified so we shall explore that now. Whenever $q=q'$ (when there is no observable transition) we have, by definition, that the probability of outcome $q$ is unity. This suggests the normalization condition
\begin{eqnarray}
\label{eqn:orthonormality} <q|q> & = & \sum_{Q} |<Q|q>|^2 \ = \ 1
\end{eqnarray} 
and, in the light of the state proximity principle for small changes and the posterior probability principle, it is natural to interpret the square availability of $Q$ given state $q$ {\em at any moment} as the probability of finding quantum numbers $Q$ given composite quantum numbers $q$ (and vice versa). This suggests that we should always interpret the squared \availability\ $|<q'|q>|^2$ as a {\em relative} probability of possible observation of a state $q'$ given that we last observed the system to be in the state $q$. However we should not, in general, interpret it as an {\em absolute} probability unless the observational context implies an observable outcome space that includes all possible states with equal detectability -- and then only when there is an unentangled clock that has ticked (since a tick on an clock that was an entangled part of the system would require $q'\ne q$). 


\subsection{A Clock-Dependent Equation Of Motion}
We can now specify our state probability, proximity and selectivity principles in terms of the probability of detecting state $r'$ at time $t(c')$ conditional on the state $r,c$ at time $t(c)$ and later detection of clock state $c'$ as
\begin{eqnarray}
\label{eqn:probDetectSensitivity} P(r',t(c');r,t(c)) & = & \frac{|s(r',c')<r',c'|r,c>|^2}{\sum_{\tilde{r}}|s(\tilde{r},c')<\tilde{r},c'|r,c>|^2}
\end{eqnarray} 
where $s(r',c')$ is the selectivity/sensitivity of the apparatus to detecting state $r'$ under the condition that the final clock state is $c'$\footnote{Intuitively, we might expect the sensitivity to depend on the \availability\ of the post-detection state of the apparatus, given its prior detection state. For the time being, we will merely note that if the apparatus is incapable of detecting certain states, then the selectivity to those states must vanish. For example, if we expect a pion in the posterior state, but our apparatus is only capable of detecting charged particles, then the selectivity for states with a $\pi^0$ must vanish.}.

If the possible final states $\tilde{r}$ are all equally detectable, then we can omit the selectivity factors $s(r,c')$ (since they are independent of $r$ and cancel out):
\begin{eqnarray}
\label{eqn:unbiasedProb} \tilde{P}(r',t(c');r,t(c)) & = & \frac{|<r',c'|r,c>|^2}{\sum_{\tilde{r}}|<\tilde{r},c'|r,c>|^2}
\end{eqnarray} 
And we see that the probability $\tilde{P}$ calculated by \ref{eqn:unbiasedProb} is not, in general, the probability of actual physical detection in a selective experiment, but the probability of outcome only for a potential {\em comprehensive unbiased} observer in our model picture of reality.

Of particular importance here is to note that
\begin{enumerate}
  \item {Whenever the prior and posterior states are identical ($r',c' = r,c$) then we have no transition and so no evolution. In order to observe a transition we must record a tick on our clock, $c' \ne c$, and so we must have $s(r',c)=0$ in eqn. \ref{eqn:probDetectSensitivity} for all $r'\ne r$.}
  \item {The squared \availability\ gives only the {\em relative} probability of finding the  {\em equally detectable} states {\em conditional on the last known state}. But it is the \availabilities\ that give the fundamental properties of a QM system, rather than experiment-dependent probabilities. They are properties of the initial set-up only and not in any way dependent on the final state detectability -- which injects a selectivity factor. Furthermore, unlike in conventional time-dependent QM, where the superposition is assumed to describe the unobserved state, our \availabilities\ apply only to a {\em possible future observation}. They say nothing about the unobserved state other than what {\em could} be observed.}
  \item {The relationship of the \availability\ to the probability of detection (or theoretical probability of outcome) includes our principle \ref{prin:proximity} (state proximity) and is a QM alternative to classical local causality in that small state changes are more likely over shorter time periods. But our principle is looser (in that it is probabilistic rather than exact) whilst simultaneously more comprehensive (in that it gives different forms of expression for different choices of clock and different kinds of state space) whilst continuing to forbid action-at-a-distance\footnote{This is implied by finite change of observables. See also the discussion in the section \ref{sec:chooseClocks}.}.}
  \item {Since the \availability\ is symmetric between the states $r,c$ and $r',c'$:
\begin{eqnarray}
|<r,c|r',c'>| \ = \ |<r',c'|r,c>|
\end{eqnarray}
then our definition of \availability\ implies {\em \availability -reversal symmetry}. This is analogous to time-reversal symmetry but not identical since the latter depends on a given choice of $t(c)$. From this point of view we attribute the observed violation of time-reversal symmetry in weak interactions to the conventional Lorentzian choice of time function $t(c)$, whereas \availability -reversal symmetry is preserved.}
\end{enumerate} 

\subsection{Isolated States And Observers}
There is much discussion in the literature and the historical development of QM concerning the concept of an isolated state and the role of the observer. Some actors in this discussion are concerned with the entanglement of any system with its environment and, in particular, with the observer. We shall concern ourselves here only with situations where we consider it a reasonable approximation to assume that a physical system may be treated as isolated from other physical systems between observations. 

In general, two entangled systems will have some composite quantum numbers preferred with individual quantum numbers in a mixed state; isolated systems must be in externally-observable (composite) eigenstates. Initial entanglement or isolation is therefore determined by the initial states -- which are part of the experimental set-up. Traditionally we assume that the separation of observer from the observed system and the assumption of isolation of the observed system are implicit in the selectivity of the observational framework. Whenever these conditions are not satisfied, then the observational framework will simply be inappropriate to non-compliant events. For example, we note that for any event in which the composite quantum numbers of the observed system are not conserved (violate eqn \ref{eqn:orthogonal}) then the event must be non-compliant with system isolation because of interaction with the environment.

A classic example of system isolation would be a particle scattering event. Although the detection of a scattered particle implies an interaction (and therefore future entanglement) with the detection equipment, we can assume that this takes place subsequent to the interaction which produced the particle (which interaction is governed by the proximity principle) and therefore it is reasonable to consider the scattering interaction itself as being confined to a system that is sufficiently isolated from the observer. 

Between observations, the internal structure of the system is not known. However, if it is isolated, then we shall assume that the {\em composite} quantum numbers  $Q$ of the system are conserved as expressed in the orthogonality relation \ref{eqn:orthogonal}. As a practical matter, we should require such conservation rules to be satisfied as an essential requirement for the isolation assumption to be valid.

Given such an isolated system, it is then that we expect \ref{eqn:qScalarProd} and \ref{eqn:probDetectSensitivity} to hold. Whenever the clock is itself isolated from the rest of the system (but possibly entangled with the observer) we can factorize the \availability\ as
\begin{eqnarray}
\label{eqn:factorClock} |<r',c'|r,c>| & = & |<r'|r>|\ |<c'|c>|
\end{eqnarray} 
and we write
\begin{eqnarray}
\label{eqn:probClockIndep} P(r',t(c');r,t(c)) & = & \frac{|s(r',c')<r'|r>|^2}{\sum_{\tilde{r}}|s(\tilde{r},c')<\tilde{r}|r>|^2}
\end{eqnarray} 
and once again we can write
\begin{eqnarray}
\label{eqn:factoredRest} <r'|r> & = & \sum_{R} <r'|R><R|r>
\end{eqnarray}
where $R$ are the composite quantum numbers of the system that starts out in state $r$ and we see that $<r'|r>$ is the well-known S-matrix element for the transition $r'\rightarrow r>$. Clearly, if the initial state is known only by its external composite variables (as, for instance, in particle decay), $r=R$, then $<r'|r> = <r'|R>$ is a vector-coupling coefficient, $R$ must be conserved and the components of the final system $r'$ must be entangled with composite quantum numbers $R$.

Also, if, again, there is no transition then $r'=r$. But suppose we observe $r'\ne r$. Then we know that there must have been a transition. In this case, we can choose the transition $r\rightarrow r'$ itself to be the clock tick and we find
\begin{eqnarray}
\label{eqn:transitionProbClockIndep} P(r';r)|_{r'\ne r} & = & \frac{|\bar{s}(r')<r'|r>|^2}{\sum_{\tilde{r}\ne r}|\bar{s}(\tilde{r})<\tilde{r}|r>|^2}\\
\label{eqn:transitionBranch} \frac{P(r_1;r)}{P(r_2;r)}\Big|_{r_1,r_2\ne r} & = & \frac{|\bar{s}(r_1)<r_1|r>|^2}{|\bar{s}(r_2)<r_2|r>|^2}
\end{eqnarray} 
where now $\bar{s}(r')$ is the sensitivity to $r'$ given that $r'\ne r$. A situation where \ref{eqn:transitionProbClockIndep} and  \ref{eqn:transitionBranch} are applicable could be particle decay or scattering where  the clock is the entire isolated system itself, an event is a tick and the actual ``time'' value of the clock is irrelevant.

\subsection{A ``Free'' Particle And The Structure Of Space-Time}\label{sec:free}

Now this relationship between transition within an isolated system and conditional availability might seem paradoxical when it comes to apparently continuous variables such as position co-ordinates of a ``free'' particle. We observe it initially at position $x$ and then at position $x'$. There is a transition (except in the rest frame) but no entanglement, no apparent change in the state space and certainly no conservation of position -- and this would seem to contradict the framework for describing transition we have built in this section. But this situation is deceptive for two reasons. First, in order to move, an isolated object must have conserved momentum and, for reasons previously explained (and well established in the uncertainty principle) this is not a compatible representation with the position representation. Second, the position or momentum of the particle is measured relative to the observer and in this case the only observed transition is in this {\em relative} position or {\em separation}. In the rest frame of the particle its position is indeed conserved.

Hence to describe this transition we cannot escape the fact of the entanglement between particle and observer. The role of the observer is {\em physically essential} since the relation between observer and particle is the only source of an observable transition. Without considering the observer we might be tempted to assume the availability is $|<x'|x>| = \delta(x'- x)$ thus forbidding any motion. It is the observation that breaks this relationship when the observer is not in the rest frame of the particle. This entanglement between observer and the observed then brings into question how the observed particle can be considered ``free''. Clearly the concept of ``free'', in the sense of unentangled, only applies to lack of entanglement with other observables, not with the observer. 

We can gain some insight into the entanglement with the observer by noting that when not in the rest frame then information about the current position $x$ will take time to reach the observer depending on the current separation (and when it does, of course, the particle may have moved on). Hence we see that the ``free'' particle in any other frame than its rest frame must be entangled with the observer's clock. The correct availability of the position $x'$ at time $t(c')$ given the position $x$ at time $t(c)$ will then be $|<x',c'|x,c>|$. We can then define a ``free'' particle {\em with respect to a given clock and a given time-function}, by the constant velocity $v$ requirement that $<x',c'|x,c> = \delta(x' - x - v(t(c')-t(c))$.

Alternatively, in the general, non-rest-frame case, the availability of the position co-ordinate of a free object for an arbitrary observer can then be computed from the transformation that takes the observer from the particle's rest frame. But this presupposes a space-time for which we know how to make the transformation. Any given spatial co-ordinate then becomes a function $x(\xi)$ of the separation $\xi$ (however we measure $\xi$) in the same way that time is a function $t(c)$ of a clock variable (again however we measure it). The ``transition'' $x\rightarrow x'$ is described entirely by the transformation of the separation $\xi\rightarrow \xi'$ between observer and particle.

Just as our concept of co-ordinate time originates in change, the concept of co-ordinate space originates in such {\em separation} of observer from observed and the subsequent projection of this concept onto the separation of two observed objects (each of which can be considered the ``observer'' of the other, if desired). Of course, the structure of that co-ordinate space (and how it transforms) is then dependent on the combination of how it is measured and how those measurements depend on the observer. In particular, how we measure the separation of two objects that are separated from the observer. The underlying dynamics, if such exists, should then be considered invariant under -- and therefore independent of -- both the transformations and the choice of measurement method. 
 
Once again,  in the picture we have laid out here, we see how entanglement is the generator of the dynamics of the system and we conjecture that our concept of co-ordinate space emerges purely because of the need to allow the separation of the observer from the observed in rather the same way that we assumed co-ordinate time to emerge from the observation of change (the separation of observations).
 
Indeed we speculate that the curvature of co-ordinate space-time that is at the heart of general relativity is a manifestation of the entanglement of the observer and their clock with the observed system when expressed in a specific co-ordinate space-time framework. And the selection of this specific space-time framework is a generalization of the selection of a co-ordinate time-function $t(c)$ to a space-time co-ordinate $\{x(\xi,c),t(\xi,c)\}$. If we choose a different space-time framework then we get different space-time equations of motion but the underlying dynamics, described by \ref{eqn:probDetectSensitivity} in terms of the variables of the observer's clock $c$ and the {\em separation} of observer from the observed system $\xi$, must be unchanged.

\section{Computing State Availability}\label{sec:compute}

We saw in the last section how the probabilities of detected outcomes depend on state \availabilities\ conditional on the initial state and the sensitivity of the detection apparatus to each possible state. We must now address certain issues regarding their computation.

This conditional \availability\ depends on an initial coupling of the prior component state to a composite state followed by decomposition into a posterior component state and we wrote it as  the magnitude of $<q'|q>$ in the composite space, to be calculated via eqn. \ref{eqn:qScalarProd} in terms of the couplings of the component product spaces to the composite space.

Clearly this calculation depends entirely on the spectrum of allowed observable states in our theoretical universe and their symmetry properties. We do not claim to have such a complete theory. As we stated at the beginning, we view QM not as a theory of reality but a framework for relating our picture of reality to reality itself. Any theory of reality then sits on top of QM. But we can show, as an example, how to calculate the \availability\ in a very simple hypothetical universe of states that have only a single SU(2) property, which for the sake of convenience and familiarity we shall call isospin and its third component and denote by $i,m$.  

The SU(2) algebra tells us how to present the direct product of two state vectors $|i_1,m_1>$ and $|i_2,m_2>$ as a vector in a single Hilbert space representing the composite isospin state $I,M$. (We could then choose either component 1 or component 2 to be a clock relative to which we compute the behavior of the other component. Alternatively we could simply choose any transition to be a clock tick.) The irreducible representations of the direct product representation are then determined by the well-known ``Clebsch-Gordon'' (vector-coupling) coefficients which are order-dependent. Specifically these are written in the form $C^{i_1 i_2 I}_{m_1 m_2 M}$ and have an order-dependence for the two component states that gives the permutation property
\begin{eqnarray}
\label{eqn:CG}C^{i_1 i_2 I}_{m_1 m_2 M} & = & (-1)^{i_1+i_2 - I}C^{i_2 i_1 I}_{m_2 m_1 M}
\end{eqnarray}
But observationally, the two permutation-related component states are effectively identical -- except where they are distinguishable only by their observable properties (in this case $i,m$) -- and nature does not care about the order in which we describe them. Since we desire a unique state vector for every observable state, then the state vector of the two-component state must be permutation invariant {\em as long as the component state descriptions have no property that reflects the order in which they are combined to produce the composite state}. There is no order-dependence in our simple two-component isospin world so, unless we arbitrarily introduce an order-dependent phase -- which we choose not to do -- then our prior compositional scalar products must also be permutation symmetric and we write 
\begin{eqnarray}
\label{eqn:vcc}<IM| i_1m_1, i_2m_2> & = & <IM| i_2m_2, i_1m_1>\nonumber\\
& = & C^{i_1 i_2 I}_{m_1 m_2 M} + (-1)^{i_1+i_2 - I}C^{i_2 i_1 I}_{m_2 m_1 M}
\end{eqnarray}
(up to an irrelevant scalar constant). Clearly the \availability\ can be non-zero only for those I for which $i_1+i_2 - I$ is even. Similarly, the decompositional couplings into a state $i_3m_3, i_4m_4$ are
\begin{eqnarray}
\label{eqn:rvcc}<i_3m_3, i_4m_4|IM> & = & <IM| i_3m_3, i_4m_4>^*
\end{eqnarray}
(and, in fact, because the Clebsch-Gordon coefficients are real, we can dispense with the conjugation). 

Substituting \ref{eqn:vcc} and \ref{eqn:rvcc} into \ref{eqn:qScalarProd} then gives us the desired availabilities of the states $i_3m_3, i_4m_4$ given a prior state $i_1m_1, i_2m_2$ necessary for computing the transition probabilities for $i_1m_1, i_2m_2 \rightarrow i_3m_3, i_4m_4$ in our primitive hypothetical world.

In the more general case, where additional quantum numbers and possible additional components are observed, the contribution of each composite state will always depend on the vector coupling coefficients  that provide the composite states as irreducible representations symmetrized with respect to permutation of complete order-independent state descriptions for each pair of component states\footnote{This may appear to contradict the well-known ``Symmetrization Postulate'' which claims that states of identical particles with half-integer spin must be anti-symmetric. However as shown in \cite{YORK} this claim is actually dependent on treating a two-valued state vector as if it is single-valued which effectively implies an implicit but subtle order-dependence in state descriptions for pairs of states with half-integer spin. This hidden order-dependence results in a physical transformation (effectively a $2\pi$ rotation on the spin quantization frame of reference of one particle relative to the other) that accompanies the permutation.  When individual component states are defined in an order-independent way, then the state vectors must be permutation symmetric {\em regardless of spin}. Of course, it is perfectly fine to anti-symmetrize half-integer spin permutation for two-particle states as long as one is consistent in using the order-dependent framework. For states of three or more particles, anti-symmetrization presents problems with defining appropriately order-dependent states and it is better to employ order-independent states as defined in \cite{YORK}.}. In effect, the fact that component distinguishability lies purely in the  properties of the component states and not in any hidden ``identity'' of the components, is what requires the permutation symmetry.

\section {Choosing Clocks}\label{sec:chooseClocks}

Classically our concept of time (the time function $t(c)$ of clock variables $c$ in our framework) was intrinsically linked to the Newtonian notion of a ``free'' object as one which continued  ``in a state of uniform motion'' (i.e. had constant velocity). Although Newton assumed it to be an observer-independent absolute, this was not essential as the space-time of special relativity showed that we could maintain the identification of a free object with constant velocity {\em for each free observer} if time and space transformed synchronously (``flat'' space-time) between free observers\footnote{That is, those moving with a constant velocity relative to each other.}. This is just like taking a position co-ordinate of a freely moving object to be a clock and choosing a time-function proportional to it\footnote{The role of the speed of light is not essential to this relationship between space and time; only to the specific equations of motion that are observed to be invariant.}. 

The advent of general relativity broke this simplistic picture by destroying our picture of a flat space-time for accelerating (interacting) objects. With the advent of QM, our concepts of dynamics then split into two separate and apparently irreconcilable paths as RQFT tried to adapt classical Lagrangian theory using the space-time of special relativity.  

The framework we have laid out here for QM and, in particular, the description of a free particle in section \ref{sec:free} suggests that further revision of space-time considerations are necessary and that although we can choose a co-ordinate space-time framework in which to express our equations of motion, it is not necessary to do so. Rather it is sufficient to identify the nature of the entanglement between objects that include (1) the ``hand'' on a clock and (2) the changing separation between observer and observed.

We have seen that although we can think of time as implicit to any dynamical theory, our only knowledge of it comes from describing change in objects we call clocks. But there is nothing special about a clock. Any object $c$ for which we can suggest a function $t(c)$ can serve as a clock. Even the occurrence of an event can itself be considered a tick on a clock. All we know about time is that given a change in any system there must always be an available choice of clock tick. But it may not always be convenient to choose the same specific observable variable $c$ as a clock with the same associated time $t(c)$ prior to every event. In this respect we can consider different clocks as being ``responsible'' for different interactions. Also, the nature of the function $t(c)$ is not unique -- even in a given frame of reference. 

However, the dynamical equations of motion with which we are familiar -- Newton's second law, Schr\"odinger's equation, Maxwell's equations, the Klein-Gordon and Dirac equations and quantum ``fields'' -- all require a unique time-function (up to an arbitrary translation). If we choose a different time function, $t(c)\rightarrow \bar{t}(c)$, then we must transform the equations of motion in a complementary way in order to obtain the same observable dynamics. In other words, our equations of motion, expressed in terms of co-ordinate time $t(c)$, depend on our choice of $t(c)$. Thus we see that if the laws of physics are to be independent of such a choice, then the time-dependent equations we choose to express them are not. Rather, to re-establish such invariance we should write our equations in terms of the clock variables $c$ directly, without the time-function intermediary. This is what the dynamical equation \ref{eqn:probDetectSensitivity} seeks to do.

The usual special relativistic concept of time implies a particular choice of $t(c)$ for a given $c$ -- one which changes from observer to observer in a particular way; according to the Lorentz time dilation when observers have a constant relative velocity, for instance. However, we should point out some generalities:

\begin{enumerate}
  \item{Whenever we observe change, it must be possible to specify the change itself as a clock tick and ``time-independent'' branching ratios such as \ref{eqn:transitionBranch} apply. But since the change event itself is the tick, the implicit clock (in this case the type of event and its environment) need not be the same clock that ticked the last time we observed a change.}
  \item{In order to describe the dynamics, it is sufficient to consider only the clock $c$ without necessarily considering $t(c)$ at all other than to relate it to the classical time.} 
  \item{The choice of a spatial separation as the time on our clock -- choosing $c$ to be the spatial separation $c =\xi$ (in a single direction) and $t(\xi)$ to be linear in $\xi$ -- necessarily implies that time transforms in the same way as the spatial separation. And any object for which $x(\xi)$ is also linear in $\xi$ we obtain transformations of $\{x,t\}$ similar to those of Lorentz invariance -- though without necessarily selecting any special role for the speed of light.} 
  \item{The finiteness requirement for data implies finiteness of observable change including clocks. Finite change in $t(c)$ is a requirement for a suitable clock that times any event. If data about any object is always finite, then no object, whether a clock or not, can have infinite speed and so {\em there must be a fastest ticking clock} and, in the case of a spatial clock, a {\em fastest moving clock} for any observer.}
  \item {Choosing the fastest changing object in a system to be a clock would naturally give us the finest detail in observed changes in the rest of the system and since fast movement facilitates rapid isolation of the clock from the rest of the system, would permit the factorization of the clock from the rest of the state availability (eqn.  \ref{eqn:factorClock}) in the asymptotic limit (as required by the conventional S-matrix for instance).}
  \item {Choosing the most strongly interacting clock would inhibit its isolation. But if a clock is not a strongly-interacting device (e.g. an electromagnetic device), then it will factor out when considering pure strong interactions only. We might then expect that a clock that interacts only through weak interactions would factor out and give us an S-matrix theory of pure quantum electrodynamics and a clock that interacts only though gravity, if such exists, would factor out when considering all but gravitational interactions. Only with gravity then is it impossible to find a clock that factorizes out unless we can find pure gravitational events for which the event itself is the clock tick.}
\end{enumerate} 

Putting these properties together we see why (a) spatial clocks synchronized to electro-magnetism have been so successful if we disregard quantum gravity and accept light as the fastest form of moving energy, (b) Lorentz-invariant S-matrix theory was successful for describing strong interactions only, (c) we need to consider other time-functions or work with clock variables directly if we are to find a dynamical theory that embraces both QM and gravity and (d) depending on the nature of the interaction being studied, we must take into account the possible entanglement of either/both clock and observer with the rest of the system being observed.

\section{Summary}
We have argued that our picture of reality is built from our observation of transition. Transitions can be of two types: transformation of the internal structure of an isolated system and transformation of the observer/object relationship. And we have shown, at least in principle, how to predict the outcome probability of such a transition conditional on a known initial state. Part of a system can be identified as a clock and the dynamics of a system is then determined by the transition in other parts relative to the clock.

As a bonus, we have seen how co-ordinate space-time emerges from our fundamental assumptions concerning the separation of objects and events rather than being an essential property of reality.

\section{Epilogue}
Since writing the essential content of this paper it has come to the author's attention that previous authors \cite{PW}, \cite{ROV} have considered the role of entangled clocks as distinct from continuous (co-ordinate) time (though in a very different way) and that a recent experiment \cite{MOR} purports to verify a particular aspect of this.

\end{document}